\documentstyle[12pt]{article}

\begin{document}

\begin{titlepage}


\begin{center}
{\Large \bf Search for  new physics in $\Delta S = 2$  
two - body ($VV$, $PP$, $VP$)  decays of the  $B^-$ - meson }\\
\vspace{1cm}
{\large \bf S. Fajfer$^{a,b}$,  P. Singer$^{c}$\\}

{\it a) J. Stefan Institute, Jamova 39, P. O. Box 3000, 1001 Ljubljana,
Slovenia}
\vspace{.5cm}

{\it b)
Department of Physics, University of Ljubljana, Jadranska 19, 1000 Ljubljana,
Slovenia}
\vspace{.5cm}

{\it c) Department of Physics, Technion - Israel Institute  of Technology,
Haifa 32000, Israel}

\end{center}

\vspace{0.25cm}

\centerline{\large \bf ABSTRACT}

\vspace{0.25cm}

The $\Delta S = 2$  $b \to \bar d ss$  transition proceeds via 
the box - diagram in the Standard Model with a branching ratio 
calculated to be below $10^{-11}$, thus providing an appropriate 
testing  ground for physics beyond the Standard Model. 
We analyze the $\Delta S = 2$ two - body 
$B^- \to K^{*-} \bar K^{*0}$, 
$B^- \to K^-  \bar K^{0}$, $B^- \to K^{*-} \bar K^{0}$ and 
 $B^- \to K^{-} \bar K^{*0}$ exclusive decays which 
 are driven by the $b \to \bar d ss$ 
 transition, both in the Standard Model 
 and in several extensions of it. 
 The models considered are the minimal supersymmetric model with and 
 without ${\cal R}$ parity conservation and two Higgs doublet 
 models. All four modes are found to have a branching ratio 
 of the order of $10^{-13}$ in the Standard Model, while 
 the expected 
 branching ratio in the different extensions vary between 
 $10^{-9} - 10^{-6}$.

\vspace{0.2cm}

\end{titlepage}

The intensive search for physics beyond
the Standard Model (SM) is performed nowdays 
 in various areas of particle physics. Among these, 
 rare B meson decays are 
suggested to give good opportunities for discovering new
physics beyond SM \cite{AM1}.
Recently, it has been suggested \cite{PAUL1,PAUL2,FS} to investigate effects of new 
physics possibly arising from $b \to ss \bar d$ or 
$b \to dd \bar s$  decays.  
As shown in  Ref. \cite{PAUL1}, the $b \to ss \bar d$ transition is
mediated in the standard model  by the box-diagram and its calculation
results in a branching ratio of nearly  $10^{-11}$, the exact value
depending on the relative unknown phase between t, c contributions  
in the box.
The $b \to d d \bar s$ branching ratio is even smaller by a
factor of  $10^2$, due to the relative $|V_{td}/V_{ts}|$
factor in the amplitudes. 
In Ref. \cite{GNK} different scenarios were used 
in the analysis of the $b \to dd \bar s$ decay, which might 
be important in 
$B^{\pm} \to K \pi$ decays. 
The authors of Refs. \cite{PAUL1,PAUL2} have 
calculated the $b \to ss \bar d$ transition in various 
extensions of the SM. It appears that for certain plausible
values of the parameters, this decay
may proceed with a branching ratio
of $10^{-8} - 10^{-7}$ in the minimal supersymmetric 
standard model (MSSM) 
and in two Higgs doublet models \cite{PAUL2}. 

Thus, decays related to the $b \to s s \bar d$ transition 
which was calculated to be very rare in the Standard Model, 
provide a good opportunity for investigating 
beyond the Standard Model physics. 
In Ref. \cite{PAUL1} it was suggested that
the most suitable channels to see effects of
the $b \to s s \bar d$ transition are the 
$B^- \to K^- K^- \pi^+$ or $\bar B^0 \to K^- K^- \pi^+ \pi^+$ 
decays.  
Moreover, when one considers
supersymmetric models with ${\cal R}$-parity violating couplings, 
it turned out that
the existing bounds on the involved couplings of the 
superpotential
did not provide any constraint on the $b \to s s \bar d$
mode \cite{PAUL1}. 
Recently, the OPAL collaboration \cite{OPAL} has set  bounds on these 
couplings from the establishment of un upper limit for 
the $B^- \to K^- K^- \pi^+$ decay 
$BR(B^- \to K^- K^- \pi^+) \leq 1.3 \times 10^{-4}$. 
The  long distance effects in $B^- \to K^- K^- \pi^+$ 
decay \cite{FS} have also been estimated recently and they have been 
found to be of the order 
$10^{-12}$, comparable in size with the short - distance SM 
contribution, thus   leaving this decay "free" for the search of new physics.
Although it appears that 
$B^- \to K^- K^- \pi^+$ or $\bar B^0 \to K^- K^- \pi^+ \pi^+$ are very good candidates
to search for the $\Delta S = 2$ transitions, we investigate here 
another possibilty for the observation of the 
 $b \to s s \bar d$ transition: the two body decays of $B^-$. 

We consider  the $VV$, $VP$, $PP$ states. 
Although in principle 
two body decays would appear to be simpler to analyze, 
there is the complication of $K^0 - \bar K^0$ mixing. Hence one needs also 
a good estimate for the $b \to s \bar s d$ transitions as well. 
Nevertheless, not all the two-body states involve neutral $K's$ and we 
shall return to this point in our summary. 
First, we proceed to describe the framework used in our analysis 
in which we concentrate on MSSM, with and without ${\cal R}$ parity 
and two Higgs doublet models as possible alternatives to the SM. 

The minimal supersymmetric extension of the Standard Model  
leads to the following effective Hamiltonian describing the $b \to s s \bar d$ 
transition \cite{PAUL1,GGMS}
\begin{eqnarray}
{\cal H} &=& \tilde C_{MSSM}
(\bar s \gamma^\mu d_L ) (\bar s \gamma_\mu b_L), 
\label{1}
\end{eqnarray}
where we have denoted 
\begin{eqnarray}
\tilde C_{MSSM} & = &   -\frac{\alpha_s^2 \delta_{12}^{d*}\delta_{23}^d}{216 
m_{\tilde d}^2}[ 24 x f_6(x) + 66 \tilde f_6(x)]
\label{2}
\end{eqnarray}
with $x = m_{\tilde g}^2 / m_{\tilde d}^2$, and the functions 
$f_6(x)$ and $ \tilde f_6(x)$ are given in \cite{GGMS}. 
The couplings 
$\delta_{ij}^{d}$ parametrize the mixing between 
the down-type left-handed squarks.
At the scale of
$b$ quark mass and by taking the existing upper limits  on 
$\delta_{ij}^{d}$ from \cite{GGMS} and \cite{PAUL1} 
the coupling $\tilde C_{MSSM}$ is estimated to be 
$|\tilde C_{MSSM}|\leq 1.2 \times 10^{-9}$ $ GeV^{-2}$ for an average 
squark mass $m_{\tilde d} = 500$ $GeV$ and $x =8$, which leads to 
an inclusive branching ratio for $b \to s s \bar d$ of 
$2 \times 10^{-7}$ \cite{PAUL1}.
The corresponding factor calculated in  SM 
\cite{PAUL1} 
is found to be  
\begin{eqnarray}
C_{SM} & = &   \frac{1}{2} \lbrack \frac{G_F^2}{2 \pi^2} m_W^2 
 V_{tb} V_{ts}^* [ V_{td} V_{ts}^* f(\frac{m_W^2}{m_t^2} )
 + V_{cd} V_{cs}^* \frac{m_c^2}{m_W^2} g (\frac{m_W^2}{m_t^2},
 \frac{m_c^2}{m_W^2})] \rbrack
\label{3}
\end{eqnarray}
with $f(x)$ and $g(x,y)$ given in \cite{PAUL1}.
Taking numerical valus
from \cite{CASO}, neglecting the CKM phases,  one estimates  
$|C_{SM}| \simeq 4 \times 10^{-12}$ $ GeV^{-2}$.

The authors of \cite{PAUL1}  have also investigated 
beyond MSSM cases  by including $R$- parity 
violating  interactions. 
 The part of the superpotential which is relevant here is 
 $W = \lambda_{ijk}^\prime L_i Q_j d_k$,
 where $i, j, k$ are indices for the families and $L, Q,d$ 
 are superfields for the
 lepton doublet, the quark doublet, and the down-type quark 
 singlet, respectively.
Following notations of \cite{CR} and \cite{PAUL1} 
the tree level effective Hamiltonian 
is 
\begin{eqnarray}
{\cal H} &=& - \sum_{ n} \frac{f_{QCD}}{m_{\tilde \nu_n}^2}
[\lambda^\prime_{n32} \lambda^{\prime *}_{n21} 
(\bar s_R b_L ) (\bar s_L d_R) +  \lambda^{\prime }_{n21} 
\lambda^{\prime *}_{n32} 
(\bar s_R d_L ) (\bar s_L b_R)]. 
\label{4}
\end{eqnarray}
The QCD corrections were found to be important for this transition 
\cite{BAGGER}. For our purpose  it suffices to 
follow \cite{PAUL1} retaining the leading order QCD result 
$f_{QCD} \simeq 2$, for $m_{\tilde \nu}= 100$ $GeV$.

 Most recently an upper bound on the specific  combination of 
 couplings entering 
 (\ref{4}) has been obtained by OPAL from a search for the 
 $B^- \to K^- K^- \pi^+$ decay \cite{OPAL}  
\begin{eqnarray}
 \sum_{n} \sqrt{ 
|\lambda^\prime_{n32} \lambda^{\prime *}_{n21}|^2 
+ |\lambda^{\prime }_{n21} 
\lambda^{\prime *}_{n32}|^2 }< 10^{-4}. 
\label{5}
\end{eqnarray}
Here we take the order of magnitude, while the OPAL result is 
$5.9 \times10^{-4}$ based on a rough estimate 
$\Gamma (B^- \to K^- K^- \pi^+) \simeq 1/4~ $ 
$ \Gamma (b \to ss \bar d)$.

The decay $b \to ss \bar d$ has been investigated using 
two Higgs doublet models (THDM) as well 
\cite{PAUL2}. These authors found that 
the charged Higgs box contribution 
in MSSM is negligible. On the other hand, THDM  
involving several neutral Higgses \cite{CS} could 
have a more sizable contribution to these modes. 
The part of the effective Hamiltonian relevant in our case is 
the tree diagram exchanging 
the neutral Higgs bosons $h$ (scalar) and $A$ (pseudoscalar) 
\begin{eqnarray}
{\cal H}_{TH} &=& \frac{i}{2} \xi_{sb} \xi_{sd} (\frac{1}{m_h^2} 
(\bar s d ) (\bar s b) - \frac{1}{m_A^2}(\bar s \gamma_5 d ) 
(\bar s \gamma_5 b) ), 
\label{6}
\end{eqnarray}
with the coupling $\xi_{ij}$ defined in \cite{CS} as a Yukawa 
coupling
of the FCNC transitions $d_i \leftrightarrow d_j$. 
In our estimation
we use the bound 
$|\xi_{sb} \xi_{sd}|/m_H^2 > 10^{-10}$ $ GeV^{-2}$, 
$H = h,A$,  which was obtained in  \cite{PAUL2} by using the 
$\Delta m_K$ limit on $ \xi_{bd}/m_H$ and assuming 
$| \xi_{sb}/m_H| >10^{-3}$. \\

We proceed now to study the effect of Hamiltonians (\ref{1}), 
(\ref{4}), (\ref{6}) on the various two body $\Delta S = 2$ decays of 
charged B - mesons. 
In order to calculate the matrix elements of the operators appearing in 
 the effective Hamiltonian, we use the factorization approximation 
 \cite{ALI,WSB1,WSB2}, which requires the knowledge of the 
 matrix elements of the current operators or the density operators. 
 Here we use the standard form factor representation 
 \cite{WSB1,ALI} of the following 
 matrix elements: 
 \begin{eqnarray}
\langle P'(p') | \bar q_j\gamma^{\mu} q_i|P(p)\rangle & = &
F_1(q^2) (p^{\mu} + p'^{\mu} - \frac{m_P^2 - m_{P'}^2}{q^2}
(p^{\mu} - p'^{\mu})) \nonumber\\
&+ & F_0(q^2) \frac{m_P^2 -m_{P'}^2}{q^2}
(p^{\mu} - p'^{\mu}), 
\label{7}
\end{eqnarray}
where $F_1$ and $F_0$ contain the contribution of vector and scalar
states respectively and $q^2 = (p - p')^2$. Also, $F_1(0) = F_0(0)$
\cite{WSB1}. For these form factors, one usually assumes pole
dominance \cite{WSB1,JURE}
\begin{eqnarray}
F_1(q^2) & = & \frac{F_1(0)}{ 1 - \frac{q^2}{m_V^2}}; \enspace
F_0(q^2)  =  \frac{F_0(0)}{ 1 - \frac{q^2}{m_S^2}}
\label{8}
\end{eqnarray}
and in order to simplify, we shall take $m_V = m_S$. 
The matrix element between pseudoscalar and vector meson is 
usually decomposed \cite{WSB2} as
\begin{eqnarray}
\label{9}
&&\langle V(q,\epsilon_{V})| 
\bar q_j\gamma^{\mu}(1- \gamma_5) q_i|P(p)\rangle =
\\
&=&{2 V(Q^2)\over m_P+m_{V}}
\epsilon^{\mu\nu\alpha\beta}\epsilon_{V\nu}^* p_\alpha
q_\beta 
+ i \epsilon^*_{V}\cdot Q {2 m_{V}\over Q^2}Q^\mu \left( A_3(Q^2) - A_0(Q^2)\right)
\nonumber\\ 
&+& i(m_P+m_{V})\left[\epsilon_{V}^{\mu *} A_1(Q^2) 
-{\epsilon^*_{V} \cdot Q\over (m_P+m_{V})^2}(p+q)^{\mu}
A_2(Q^2)\right] \; \nonumber,
\end{eqnarray}
where $Q = p - q$.
\begin{equation}
\label{9a}
A_3(Q^2)-{m_H+m_V\over 2 m_V}A_1(Q^2)+
{m_H-m_V\over 2 m_V}A_2(Q^2)=0\;,
\end{equation}
and $A_3 (0) = A_0(0)$. 
For the vector and axial vector form factor we use again 
pole
dominance \cite{WSB1,JURE}, and relevant parametrs 
are taken from \cite{ALI,WSB2} $F_0^{BK}(0) = 0.38$, 
$A_0^{BK*}(0) = 0.32$. 
For the calculations of the density operators we use the relations
\begin{eqnarray}
\partial^{\alpha} (\bar s \gamma_\alpha b)& = &i (m_b - m_s) \bar s b
\label{10}
\end{eqnarray}
and
\begin{eqnarray}
\partial^{\alpha} (\bar s \gamma_\alpha \gamma_ 5 b) & = &
i (m_b + m_s) \bar s \gamma_ 5 b
\label{11}
\end{eqnarray}
We will use also the following decay constants:
\begin{equation}
\langle V(\epsilon_V,q)|  \bar q_j\gamma^{\mu} q_i |0\rangle = \epsilon_{\mu}^{*} (q) g_V(q^2)~,
\label{12} 
\end{equation}  
and 
\begin{equation}
\langle P(q)|  \bar q_j\gamma^{\mu} \gamma_ 5 q_i |0\rangle = i f_P q_\mu
\label{13} 
\end{equation} 
with  $f_K= 0.162$ $GeV$, $g_{K^*}= 0.196$  $GeV^2$ \cite{WSB2}. 
Now we turn to the analysis of the specific modes.\\

{\it a) $B^- \to K^{*-} \bar K^{*0}$ decay}\\

For the analysis of pseudoscalar meson decay to two vector mesons  
it is convenient to use helicity formalism (see e.g. \cite{EK}). 
We denote ${\cal O} =
(\bar s \gamma^\mu(1 - \gamma_5) d )$ 
$ (\bar s \gamma_\mu (1 - \gamma_5)b)$, and then we use 
${\cal H} = C {\cal O}$ with $C$ being $1/4 \tilde C_{MSSM}$,   
$1/4 C_{SM}$. 
Using factorization and the definitions given above, one finds
the following helicity amplitudes
\begin{eqnarray}
H_{00} (B^- \to K^{*-} \bar K^{*0} )
   =  C g_{K^*} (m_B + m_{K^*})[ \alpha A_1^{BK^*} (m_{K^*}^2) 
  - \beta A_2(m_{K^*}^2)]&&
  \label{14}
 \end{eqnarray} 
\begin{eqnarray}
H_{\pm \pm} (B^- \to K^{*-} \bar K^{*0} )
   =  C g_{K^*} (m_B + m_{K^*})[ \alpha A_1^{BK^*} (m_{K^*}^2) 
 \mp  \gamma  V^{BK^*} (m_{K^*}^2) ]&&
  \label{15}
 \end{eqnarray} 
where
\begin{eqnarray}
\alpha & = &  \frac{1 - 2 r^2}{2 r ^2}, \enspace 
\beta = \frac{k^2}{2 r^2 (1 + r)^2}, \enspace 
\gamma = (1- 4  r^2 )
  \label{16}
 \end{eqnarray} 
with $r = m_{K^*}/m_B$, $k^2 = 1+r^4 + t^4 - 2 r^2 $ $ - 2 t^2 - 
2 r^2 t^2$. 
The decay width is then 
\begin{eqnarray}
\Gamma (B^- \to K^{*-} \bar K^{*0} )
  & = & \frac{ |\vec{p}|}{8 \pi m_B^2} [ |H_{00}|^2 + |H_{++}|^2
+ |H_{--}|^2]. 
 \label{17}
 \end{eqnarray} 
Within MSSM model the branching ratio  
becomes $\leq 6.2 \times 10^{-9}$, while SM gives this rate to be  
$ 6.8 \times 10^{-14}$. 
The ${\cal R}$ - parity term described by the effective Hamiltonian (4) cannot be seen 
in this decay mode when factorization
approach is used,  since the density operator matrix element 
$\langle \bar K^{*0} | (\bar s d)| 0\rangle$ vanishes. The two Higgs doublet 
model also cannot be tested in this mode due to the same reason.\\

{\it b) $B^- \to K^{*-} \bar K^0$ decay}\\

The matrix element of the operator ${\cal O}$ is calculated to be
\begin{eqnarray}
\langle \bar K^0(k_0)  K^{*-}(k_- ,\epsilon) | {\cal O} |B^- (p_B)\rangle 
& = & - 2 m_{K^*} f_K A_0^{BK^*}
 (m_{K^*}^2) \epsilon^* \cdot k_0
  \label{18}
 \end{eqnarray} 
Denoting the decay amplitude by ${\cal A}$, one finds 
\begin{eqnarray}
\sum_{pol} |{\cal A}|^2 &= & |C|^2 f_K^2 |A_0 (m_{K}^2)|^2
\lambda(m_B^2,m_K^2,m_{K*}^2) 
\label{19}
 \end{eqnarray} 
with the $\lambda(a,b,c) = a^2 + b^2 + c^2 -2 (a b + b c + a c)$. 
The branching ratio is 
straightforwardly found to be 
$ BR( B^- \to K^{*-} \bar K^0)_{MSSM} \leq 1.6 \times 10^{-9}$, 
which is comparable to the SM prediction of Ref.  \cite{ALI} for the 
$\Delta S = 0$ $ B^- \to K^{*-} K^0$ decay, given as 
$ BR( B^- \to K^{*-} K^0) = 1  \times 10^{-9}$,  
$ 5 \times 10^{-9}$, $ 2 \times 10^{-9}$ obtained for the 
number of colours $N_c = 2$,  $N_c = 3$,  $N_c = \infty$, 
 respectively. 
 
The  SM calculation for the $\Delta S= 2$ transition 
leads to 
$ BR( B^- \to K^{*-} \bar K^0)_{SM} = 1.7 \times 10^{-14}$.  
The MSSM  which includes ${\cal R}$ parity breaking terms 
can occur in this decay. The matrix element of the 
operator ${\cal O}_{{\cal R}} = (\bar s (1 + \gamma_5) d )$ 
$ (\bar s (1 - \gamma_5)b)$ can be found to be
\begin{eqnarray}
 \langle\bar K^0(k_0)  K^{*-}(k_- ,\epsilon)| {\cal O}_{{\cal R}}|
 B^-(p_B)\rangle  &= &\nonumber\\
 \frac{m_K^2 f_K}{(m_s + m_d) (m_s + m_b)}
(2 m_{K*} \epsilon^* \cdot k_0 ) A_0^{BK^*}(m_K^2).&&
\label{R19}
 \end{eqnarray} 
Taking the values of the quark masses as in \cite{ALI}
$m_b = 4.88$ $GeV$, $m_s= 122$ $MeV$, $m_d = 7.6$ $MeV$ and using 
the bound given in Eq. (\ref{5}) we obtain the
estimation of the upper limit of the branching ratio 
$ BR( B^- \to K^{*-} \bar K^0)_{{\cal R}}$ to be 
 $ 4.4 \times 10^{-8}$. This limit can be raised 
 to  $1.5\times 10^{-6}$ for the upper bound on the couplings of 
  $5.9 \times 10^{-4}$ given in \cite{OPAL}. 

The two Higgs doublet model (\ref{6}) gives for the amplitude of 
this decay
\begin{eqnarray}
{\cal A}_{THDM}(B^- (p_B)\to  \bar K^0(k_0)  K^{*-}(k_- ,\epsilon)
& = & \nonumber\\
\frac{i}{2} \frac{\xi_{sb} \xi_{sd}}{m_A^2}
[ 2 m_{K^*} f_K A_0^{BK*}(m_{K}^2) \epsilon^* \cdot k_0 ] 
 \frac{m_K^2 f_K}{(m_s + m_d) (m_s + m_b)},&&
\label{THDM19}
 \end{eqnarray} 
which gives for the limit $|\xi_{sb} \xi_{sd}|/m_H^2 > 10^{-10}$ $ GeV^{-2}$,
a branching ratio of the order $10^{-11}$. Due to
specific combination of the products of the scalar (pseudoscalar) 
densities this is the only decay which has nonvanishing amplitude 
within the factorization assumption.\\

{\it c) $B^- \to K^{-} \bar K^{*0}$ decay}\\

For this decay mode the matrix element of the operator ${\cal O}$ 
is determined  to be
\begin{eqnarray}
\langle\bar K^{*0} (k_0, \epsilon))  K^{-}(k_-) | {\cal O} |B^- (p_B)\rangle 
& = &  2 g_{K^*} f_K F_1^{BK*}
 (m_{K*}^2) \epsilon^* \cdot k_-
  \label{18a}
 \end{eqnarray} 
giving the branching ratio in MSSM with an upper limit 
\begin{equation}
BR(B^- \to K^{-} \bar K^{*0})_{MSSM} = 5.9 \times 10^{-9}
\label{19a}
\end{equation}
in comparison with SM result $6.5 \times 10^{-14}$.

The amplitude calculated in MSSM including ${\cal R}$ breaking and 
THDM vanishes, due to vanishing of the matrix element of the 
density operator for $\bar K^{*0}$
state.\\

{\it d) $B^- \to K^{-} \bar K^0$ decay}\\

The matrix element of the 
operator ${\cal O}$ becomes in this case 
\begin{eqnarray}
\langle\bar K^0(k_0)  K^{-}(k_-) | {\cal O} |B^- (p_B)\rangle & = & 
i f_K F_0^{BK}(m_K^2)(m_B^2 - m_K^2).
\label{20}
\end{eqnarray} 
The multiplication with the corresponding 
$1/4 \tilde C_{MSSM}$ gives the  required amplitude 
${\tilde{\cal A}}$.
The branching ratio is then 
\begin{eqnarray}
\Gamma (B^- \to K^- \bar K^0) & = & \frac{1}{16 \pi m_B^2}
\sqrt{ m_B^2 - 4 m_K^2} |{\tilde{\cal A}}|^2, 
\label{21}
 \end{eqnarray}  
The branching ratio for MSSM is found to be 
$BR(B^- \to K^- \bar K^0)_{MSSM} \leq 2.3 \times 10^{-9}$,
in comparison with the $2.5 \times 10^{-14}$ found in the SM. 
The matrix element of the $R$ parity breaking MSSM operator 
 $ {\cal O}^{(1)} =   (\bar s \gamma_5 d)$$( \bar s b )$  
 is found to be 
 \begin{eqnarray}
\langle K^- \bar K^0| {\cal O}^{(1)} |B^-\rangle& =&
\langle  \bar K^0| \bar s \gamma_5 d|0\rangle
\langle K^- | \bar s b |B^-\rangle \nonumber\\ 
&  =& - i \frac{m_K^2}{(m_s + m_d)(m_b + m_s)} f_K F_0^{BK} (m_K^2)
(m_B^2 - m_K^2)\nonumber\\
\label{22}
 \end{eqnarray}  
while the operator $(\bar s \gamma_5 b) ( \bar s d)$  gives the same result 
with the opposite sign.  
The decay width is then
\begin{eqnarray}
\Gamma (B^- \to K^- \bar K^0)_{{\cal R}} & = & 
\frac{1}{16 \pi m_B^2}
\sqrt{ m_B^2 - 4 m_K^2} |\langle K^- \bar K^0| {\cal O}^{(1)} |B^-\rangle/4|^2 
\nonumber\\
&\times&\frac{f_{QCD}^2}{m_{\tilde \nu}^4} 
(\sum_{i = n}  
|\lambda^\prime_{n32} \lambda^{\prime *}_{n21}|^2 
+ |\lambda^{\prime }_{n21} 
\lambda^{\prime *}_{n32}|^2 )
\label{23}
 \end{eqnarray}  
The  constraint in (\ref{5}) gives the bound 
$9.4 \times 10^{-8}$, while 
for the bound of $5.9 \times 10^{-4}$ for the coupling constants 
(\ref{6}) the rate 
$BR(B^- \to K^- \bar K^0)_{{\cal R}}$ can reach  
$3.3 \times 10^{-6}$.

 The long distance effects are usually suppressed in the $B$ meson 
decays. One might wonder if they are important in  
decays we consider here. We have estimated the tree level 
contribution of the 
$D (D^*)$ which then goes into $K (K^*)$ via weak annihilation. 
We found that these  contributions give a branching ratio 
of the order $10^{-18}$ and therefore they can be safely 
neglected.
One might think that  the exchange of two intermediate states 
$D (D^*)$, $K (K^*)$  can introduce certain long distance 
contributions.
In decay $B \to "D"~ "K" \to "K"~"K"$ 
the first weak vertex arises from the decay  $B \to "D" ~"K"$ 
and the second weak 
vertex (see e.g. \cite{FS}) can be generally obtained
from the three body decays of $D\to KKK$. 
In Ref.  \cite{FS} it was found that such contributions are 
also very small. 
Therefore, we are quite confident to suggest that the long distance effects 
are not important in the two body $\Delta S = 2$ $B$ decays.\\ 

Let us turn now to the possibility of detecting these decay modes. 
The $B^- \to K^{*-} \bar K^{*0}$ and 
$B^- \to K^- \bar K^{*0}$ modes 
have clean signatures of a $\Delta S =2$ transition and therefore 
these are the channels we recommend to look for. The other two modes we 
discussed, $b)$ and $d)$ have a $\bar K^0$ in the final states which 
complicates the possibilty of a detection because of $K^0$  
$- \bar K^0$ mixing. Separating the desired amplitude 
requires the measurement of the decays of both $K_S$ and $K_L$, 
since one can express  \cite{BY}
\begin{eqnarray}
\frac{\Gamma (B^- \to K^- K_S) - \Gamma (B^- \to K^- K_L)}{  
\Gamma (B^- \to K^- K_S) + \Gamma (B^- \to K^- K_L)}& = & 
Re \eta(B^- \to K K^-), 
\label{24}
 \end{eqnarray}  
where 
\begin{eqnarray}
Re \eta(B^- \to K K^-) & = & \frac{A(B \to \bar K^0 K^-)}{
A(B \to K^0 K^-)}. 
\label{25}
 \end{eqnarray}  

We summarize our results in the Table 1. 
The MSSM gives rates of the
order $10^{-9}- 10^{-8}$, while  the ${\cal R}$ 
parity breaking terms in the MSSM can be seen only in the  
$B^- \to K^{*-} \bar K^{0}$ and $B^- \to K^{-} \bar K^{0}$ decay.
These are the modes which as we mentioned are more difficult 
on the experimental side. 
The THDM model can give nonvanishing contribution 
only in the case of $B^- \to K^{*-} \bar K^{0}$ decay, 
with a rate too small to be seen. Thus, we conclude by stressing the possibilty 
of detecting physics beyond SM mainly in 
the $K^{*-} \bar K^{*0}$, $K^{-} \bar K^{*0}$ decays. \\

We thank Y. Rozen, S. Tarem and D. Zavrtanik for stimulating 
discussions on experimental aspects of this investigation.\\

This work has been supported in part by the Ministry of
Science of the Republic of Slovenia (SF) and by the Fund for Promotion
of Research at the Technion (PS).\\

\begin{table}[h]
\begin{center}
\begin{tabular}{|c||c|c|c|c|}
\hline
 Decay & SM & MSSM & ${\rm MSSM} + {\cal R}$ & THDM \\
\hline
$B^- \to K^{*-} \bar K^{*0}\phantom{\Big|}$ & $ 6.9 \times 10^{-14}$ &
$ 6.2 \times 10^{-9}$ & $ - $ & $ -$\\
$B^- \to K^{*-} \bar K^{0}$ & $ 1.7 \times 10^{-14}$ &
$ 1.6\times 10^{-9}$ & $   10^{-7} - 10^{-6}$ & $ 10^{-11}$\\
$B^- \to K^{-} \bar K^{*0}$ & $ 6.6 \times 10^{-14}$ &
$ 5.9\times 10^{-9}$ & $ - $ & $ - $\\
$B^- \to K^{-} \bar K^{0}$ & $ 2.5 \times 10^{-14}$ &
$ 2.3 \times 10^{-9}$ & $ 10^{-7} -10^{-6}$  & $ - $\\
\hline
\end{tabular}
\caption{ The predicted branching ratio for the 
$B^-$ $\Delta S= 2$ two - body decays calculated using the 
factorization approach within  Standard Model (the first column), 
Minimal Supersymmetric Standard Model (the second column), 
Minimal Supersymmetric Standard Model extended  by 
${\cal R}$ parity breaking (the third column), and Two Higgs Doublet Model
(the fourth column). The values in columns two, three, four 
are upper limits, as determined from present knowledge of upper 
limits for couplings involved. } 
\end{center}
\end{table}

\end{document}